# In vacuo synthesis of single-layer $Ni_3(HITP)_2$ on HOPG surface using metallo-organic precursor

Chuyu Song†, Yifei Feng†, Henan Chen, Nian Lin*

Department of Physics, The Hong Kong University of Science and Technology, Clear Water Bay, Hong Kong 999077, China

†These authors contributed equally to this work

*Corresponding author: phnlin@ust.hk

Abstract

Single-layer conjugated metal–organic frameworks (SL c-MOFs) are predicated theoretically to host rich quantum phases. To date, however, the experimental synthesis has largely resulted in SL c-MOFs on metal substrates, whose intrinsic properties are strongly screened due to substrate hybridization. To overcome this obstacle, here we develop a method to grow clean SL c-MOF $Ni_3(HITP)_2$ on a chemically inert substrate of highly oriented pyrolytic graphite (HOPG). By means of an ultra-high vacuum based multi-step protocol using nickel acetylacetonate [$Ni(acac)_2$] precursor, we obtain continuous single-layer $Ni_3(HITP)_2$. Scanning tunneling microscopy reveals a well-defined hexagonal framework with a uniform azimuthal orientation rotated by approximately 30° with respect to the underlying graphite lattice. Furthermore, we control the growth of bilayer $Ni_3(HITP)_2$ which features an unusual AA stacking configuration. This work establishes metallo-organic chemistry as an efficient route for integrating 2D c-MOFs onto inert substrates, which opens a new avenue for exploring the exotic quantum properties of SL c-MOFs.

## Introduction

Metal–organic frameworks (MOFs) are porous crystalline materials constructed from metal nodes and organic linkers [1,2]. Owing to exceptional structural tunability and versatile functionalities, MOFs have been studied extensively in the past three decades [3-5]. Among them, two-dimensional conjugated MOFs (2D c-MOFs) represent a unique type of 2D materials [6-14]. Specifically, theoretical predictions have envisioned that rich quantum phases can be realized in 2D c-MOFs such as quantum spin Hall state, quantum anomalous Hall state, Chern flat band, quantum spin liquids, etc.[15-26]. Most 2D c-MOF sample are in the form of thin films or multilayers with their thickness ranging from 1~2 μm to 10 nm. Single layer (SL) 2D MOFs synthesized by the means of conventional "wet" chemistry are rarely reported. [27]. Considering the quantum phases are prone to inter-layer interactions, single-layer (SL) c-MOF samples are highly desirable. In this regard, on-surface coordination assembly on metal substrates offers an efficient approach for preparing SL-2D MOFs [28,29]. However, the strong screening of the metal substrate can obscure the intrinsic properties of the MOF.

Synthesizing SL c-MOFs on chemically inert substrates like graphene [30-32], h-BN [33,34], or other van der Waals materials [35] offers an effective approach to minimize substrate screening effects and preserve their intrinsic properties. Our previous studies have achieved SL c-MOFs on inert substrates of HOPG and MoS2 [36,37]. Nevertheless, these systems are based on linker molecules that do not require dehydrogenation because dehydrogenation reactions on the inert substrates remain a significant challenge. In addition, metal atoms deposited on the inert substrates tend to aggregate into large clusters rather than dispersed as single atoms, which complicates the formation of large crystalline domains.

Here we demonstrate a strategy utilizing metallo-organic precursors of metal acetylacetonate complexes [M(acac)x] to synthesize SL c-MOFs on chemically inert substrates. The acetylacetonate ligand offers several distinct advantages [39-41]: (i) its strong bidentate chelation stabilizes the metal center in a +2 oxidation state, preventing metal aggregation or reduction in the course of sublimation and deposition; (ii) the acac moiety undergoes clean thermal decomposition at moderate temperatures, producing only volatile byproducts (acetone, $CO_2$) with no residue; and (iii) $M(acac)_x$ compounds can be sublimated at moderate temperatures, allowing for precise control over metal precursor delivery. These characteristics make acetylacetonates suitable for forming SL c-MOF on inert substrates.

$Ni_3(HITP)_2$ is predicted to exhibit non-trivial quantum phase [42-44]. In this work, we demonstrate that $Ni(acac)_2$ enables controlled growth of single-layer $Ni_3(HITP)_2$ on an HOPG substrate through a multi-step protocol. The resulting $Ni_3(HITP)_2$ single layers exhibit good crystallinity with a homogenous orientation with respect to the graphite lattice. We also achieve controlled synthesis of a bilayer structure and identify an AA-stacked bilayer. We test the stability of the $Ni_3(HITP)_2$ single layers and find that they retain the structural integrity after 350 °C heating and air exposure, making them accessible for device integration.

**Results and discussion**

The multi-step growth protocol is illustrated in Figure 1. 2,3,6,7,10,11-hexaiminotriphenylene (HATP) linker molecules and $Ni(acac)_2$ precursors are co-deposited onto an HOPG substrate held at the room temperature (i). The substrate is then annealed at ~250 °C for 10 minutes (ii). At this temperature, the $Ni(acac)_2$ precursors are fragmented to release $Ni^{2+}$ ions and acac ligands. The acac ligands capture protons from the amino groups of HATP to form neutral Hacac molecules that desorb from the HOPG surface. The deprotonation converts HATP into its reactive form 2,3,6,7,10,11-hexaiminotriphenylene (HITP), which coordinate with the $Ni^{2+}$ ions, resulting in single-layer $Ni_3(HITP)_2$ domains on the substrate (iii). The un-reacted $Ni(acac)_2$ precursors and the partially fragmented precursors aggregate as clusters, denoted as $Ni(acac)_x$ clusters. Figure 2a shows the representative sample morphology of iii. One can see the hexagonal framework of the single-layer $Ni_3(HITP)_2$ formed on the left part of the image and three large clusters and two small cluster are at the edges of the $Ni_3(HITP)_2$ domains. The apparent size of the clusters is in the range of 6~7 nm laterally and 2.5~3.5 nm vertically (Figure S1a).

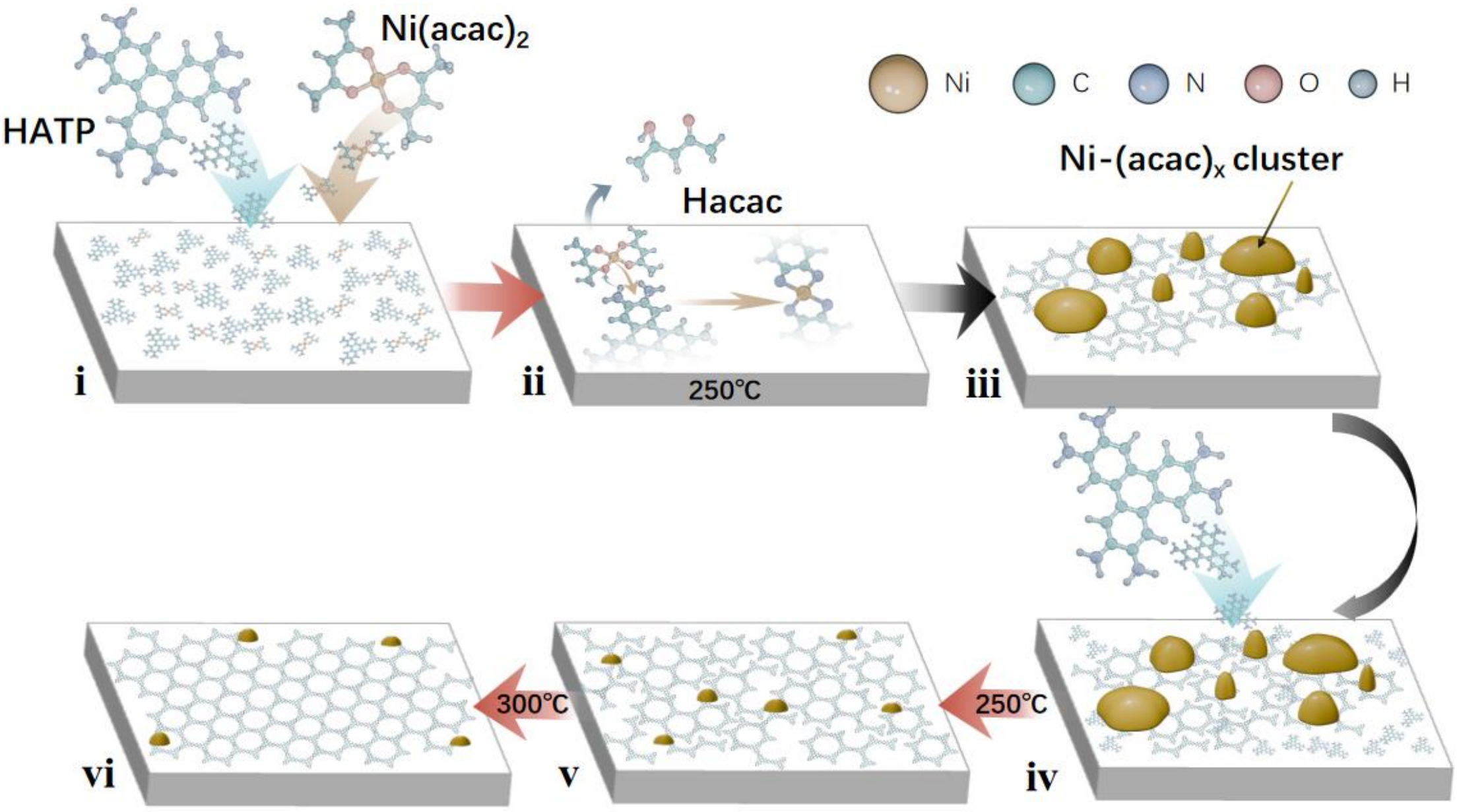

Figure 1. Schematic Illustration showing the multi-step growth process: (i) Co-deposition of HATP and $Ni(acac)_2$ precursors on HOPG; (ii) On-surface HATP deportation and $Ni^{2+}$ coordination; (iii) Formation of $Ni_3(HITP)_2$ single layer together with $Ni(acac)_x$ clusters; (iv) Cycles of subsequential reaction with additional HATP; (v) Formation of single layer $Ni_3(HITP)_2$ composed of small domains; (vi) Formation of single layer $Ni_3(HITP)_2$ composed of large domains.

In the next step (iv), additional HATP are deposited on the sample, which is annealed at ~250 °C for 10 minutes. This process is repeated several times. As a result of such cycles, more and more single-layer $Ni_3(HITP)_2$ domains are formed while the $Ni(acac)_x$ clusters are gradually reduced in size (5.0~5.5 nm laterally and 1.5~2.5 nm vertically) and in density (Figure S1b) (v). We propose, in these cycles, the $Ni(acac)_x$ clusters provide acac ligands to deprotonate the later supplied HATP molecules and $Ni^{2+}$ ions to coordinate with the resultant HITP to form $Ni_3(HITP)_2$. Presumably, the repeating reactions recycle the unreacted and partially fragmented $Ni(acac)_2$ precursors.

When the $Ni\text{-}acac_x$ clusters are nearly consumed, the single-layer $Ni_3(HITP)_2$ covers ~80% of the surface (Figure 2b). Closer inspections reveal that the single layer is composed of small domains of ~10 nm size. These domains are in the same azimuthal orientation (Figure S2) and separated by misalignment domain boundaries and defect lines. Presumably, different domains are formed out of different nucleation seeds in the growth process. To enlarge domain size, the sample is annealed at ~300 °C for 10 minutes (vi). Figure 2c shows that the resultant single layer exhibits enlarged domains. The domains can reach a size of ~30 nm, are more closely packed and better ordered. The misalignment domain boundaries and defect lines are reduced significantly.

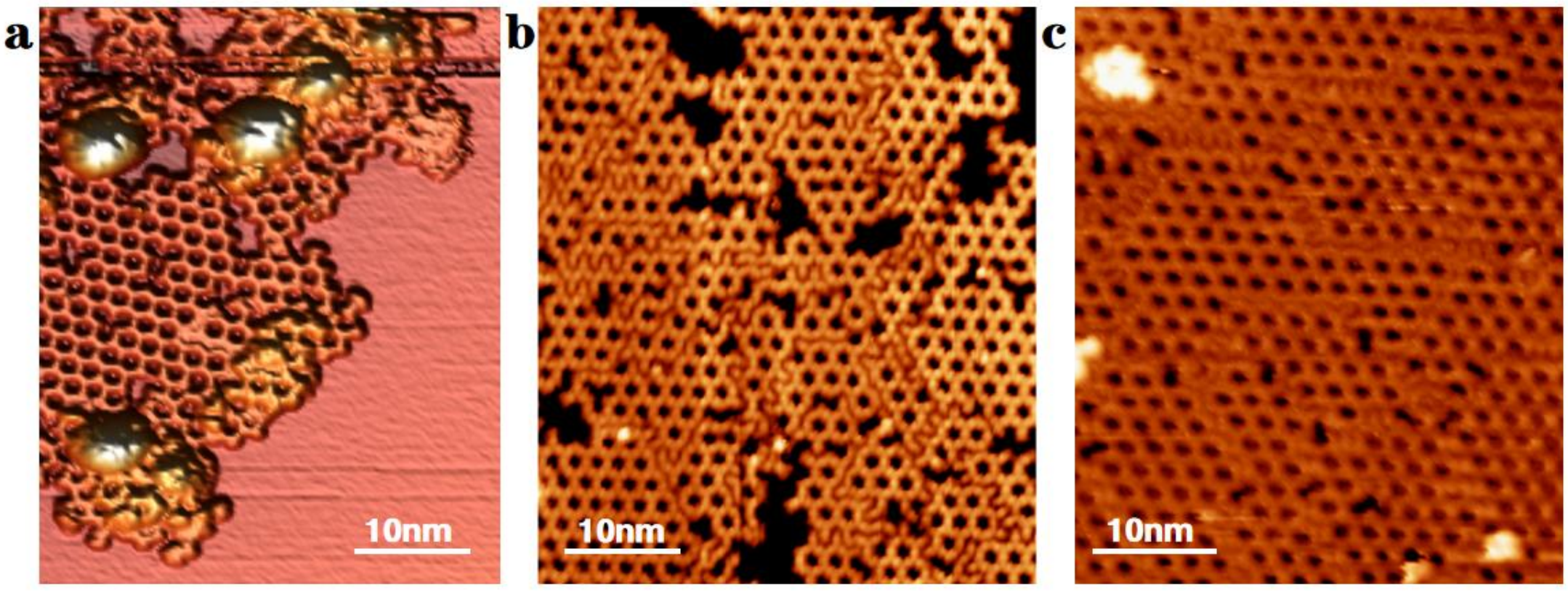

Figure 2. STM images showing the sample morphology of steps (iii), (v) and (vi) in the multi-step growth. (a) Single-layer $Ni_3(HITP)_2$ domains with $Ni(acac)_x$ clusters. (b) Single-layer $Ni_3(HITP)_2$ composed of small domains. (c) Single-layer $Ni_3(HITP)_2$ composed of large domains. Scanning parameters: $V = -2$ V, $I = 500$ pA.

We conducted series of controlled experiments to gain further insights into the growth mechanism. First, we co-deposited metallic nickel and HATP on the HOPG substrate held at room and elevated temperatures, respectively. Neither method yields $Ni_3(HITP)_2$. Ni atoms form large clusters adhered at the steps of the HOPG substrate (Figure S3a), in contrast, $Ni(acac)_2$ aggregate as small clusters on the open terraces of the HOPG substrate (Figure S3b). Presumably, the Ni atom clusters do not provide reactive agents to deprotonate HATP. Secondly, we co-deposited HATP and $Ni(acac)_2$ simultaneously on a HOPG held at high temperature (~300 °C). This method yields small pieces of $Ni_3(HITP)_2$ at step edges of the HOPG substrate (Figure S4). This result suggests that the co-deposited $Ni(acac)_2$ and HATP at the hot substrate easily desorb from the substrate, which is unfavored for growing large $Ni_3(HITP)_2$ domains. Finally, to test the substrate effects, we replace HOPG with a single crystal Au(111) as the substrate. Unlike those formed on HOPG, small domains of $Ni_3(HITP)_2$ that are scattered and orientated in multiple directions are formed on the Au(111) substrate (Figure S5).

Figure 3a is a high-resolution STM image of the $Ni_3(HITP)_2$ single layer, displaying a hexagonal framework. The red rhombus frame highlights the unit cell of $Ni_3(HITP)_2$. The lattice constant of the unit cell is 2.19±0.02 nm, which is consistent with the free-standing $Ni_3(HITP)_2$ lattice constant [42,45]. The Ni atoms could not be resolved. This is attributed to charge delocalization due to strong conjugation of $Ni_3(HITP)_2$. The blue arrows indicate $a_1$ and $a_2$ crystallographic directions of the substrate graphite lattice, indicating that the $Ni_3(HITP)_2$ lattice is rotated by 30° with respect to the underneath graphite atomic lattice. Figure 3b depicts a molecular model of the $Ni_3(HITP)_2$ single layer on the graphite substrate. Note that the registration of $Ni_3(HITP)_2$ on the graphite atomic lattice is tentative since the underneath graphite atoms could not be resolved simultaneously.

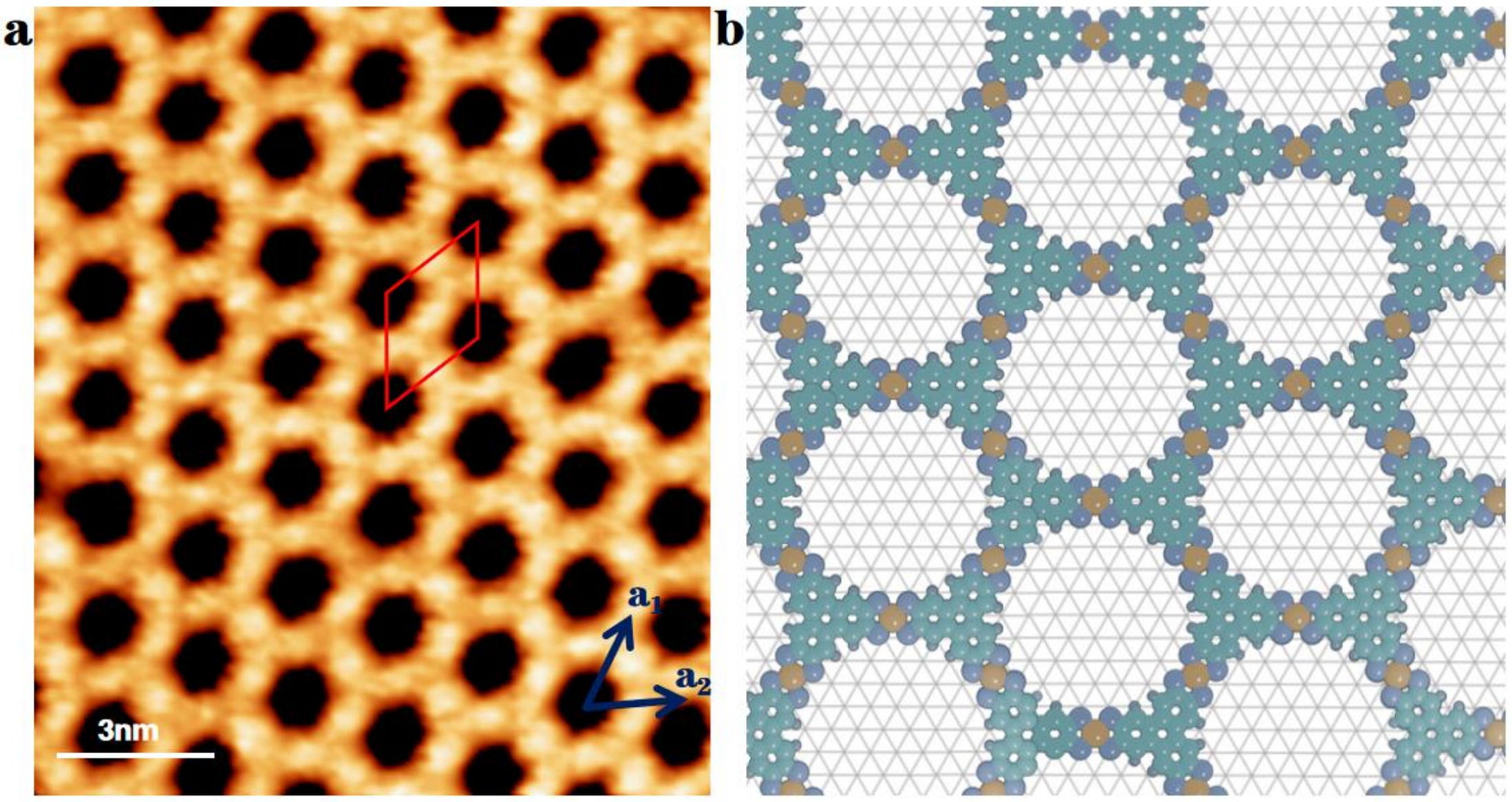

Figure 3. $Ni_3(HITP)_2$ single layer. (a) High-resolution STM image. The red rhombus frame highlights the unit cell and the blue arrows indicate $a_1$ and $a_2$ crystallographic directions of the underlying graphite atomic lattice. Scanning parameters: V = -2 V, I = 500 pA. (b) Tentative molecular model of the $Ni_3(HITP)_2$ single layer adsorbed on the graphite lattice.

After forming the $Ni_3(HITP)_2$ single layer, depositing excessive HATP to react with the remaining $Ni(acac)_x$ clusters results in second layer $Ni_3(HITP)_2$ formed on top of the single layer. Figure 4a shows the two layers exhibit the same lattice orientation and lattice constant. Figure 4c is the cross-sectional height profile of line ABC in Figure 4a. In section AB, the peak-valley corrugation of the second layer aligns with that of the first layer as guided with the blue rulers, indicating an AA stacking configuration of the bilayer structure, as illustrated in Figure 4b. In section BC, the cross-sectional height profile reveals that the first layer is ~0.40 nm above the HOPG substrate, and the second layer is ~0.35 nm above the first layer. This value is a typical interlayer distance of van der Waals layer materials. Figure 4d shows a side view model of the bilayer structure.

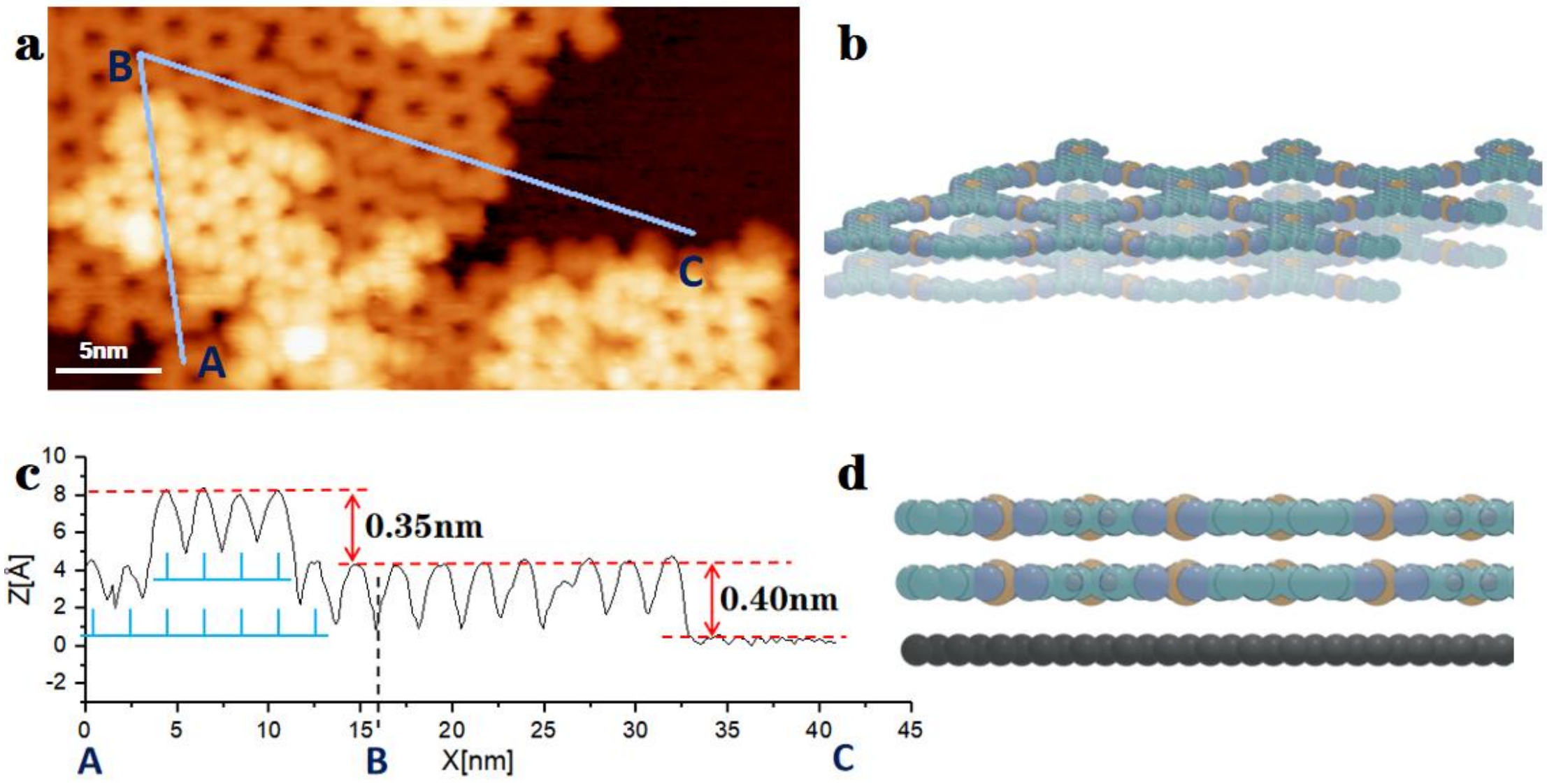

Figure 4. Bilayer $Ni_3(HITP)_2$ structure. (a) STM image. Scanning parameters: V = −2 V, I = 500 pA. (b) AA stacking of the bilayer structure. (c) Cross-sectional height profile of ABC in (a), showing the AA-stacking alignment of the two layers and the inter-layer distance. (d) Side view model of the bilayer structure on graphite.

Last but not the least, we tested the structural stability of the $Ni_3(HITP)_2$ single layer. We found that after annealing at 350 °C for 10 min, the framework structure remained intact (Figure S6a). We took the sample out of vacuum and exposed it to the ambient air for 30 minutes. After introducing the sample to vacuum and annealing at 250 °C to remove the ambient adsorbates, the framework structure remained intact (Figure S6b). These results suggest that the $Ni_3(HITP)_2$ single layer is suitable for further processing and device integration.

## Conclusion

In summary, we have developed an effective method of using metallo-organic precursor to grow clean single layers c-MOF on an inert substrate. The multi-step growth protocol yields continuous c-MOF single layers with a uniform crystallographic orientation with respect to the underlying substrate lattice. This method highlights two key ingredients for forming c-MOF on the inert substrate, namely, preventing metal aggregation and activating ligand deprotonation. We anticipate that this method is applicable for other metallo-organic precursors and other linker molecules. The bilayer growth offers promising prospect of using different precursors subsequentially to grow layer-by-layer heterostructure c-MOFs. Considering the intrinsic properties of c-MOFs grown on inert substrates are preserved, these samples are suitable for fabricating c-MOF based devices.

## Methods

All measurements were conducted using an Omicron ultrahigh vacuum (UHV) scanning tunneling microscope (VT-STM) system with a base pressure of $1\times10^{-10}$ mbar. All STM images were acquired in constant-current mode at the room temperature. HOPG substrates were prepared by cleaving with adhesive tape followed by annealing at approximately 330 °C in the UHV chamber for 3 hours. HATP molecules (purchased from CHEMSOON, purity: 95%) were sublimed at 310 °C using an organic-beam evaporator. $Ni(acac)_2$ precursors (purchased from Bio-Dream, purity: 96%) were sublimed at 60 °C using the same method. The integrity of the $Ni(acac)_2$ precursors was checked with deposition on a Au(111) substrate (Figure S7).

## Acknowledgement

The work was supported by the Hong Kong RGC 16300425.

## Author Contributions

C. S and Y. F. contributed to designing the method, conducting the experiments, acquiring and analyzing the data, and preparing the manuscript. H. C. contributed to designing the method. C. S and N. L contributed to writing the manuscript. N. L. supervised the project. All authors have given approval to the final version of the manuscript.

## Competing interests

The authors declare no competing interests.

## Data Availability Statement

The datasets generated during and/or analysed during the current study are available from the corresponding author on reasonable request.

Supporting information:

# In vacuo synthesis of single-layer $Ni_3(HITP)_2$ on HOPG surface using metallo-organic precursor

Chuyu Song†, Yifei Feng†, Henan Chen, Nian Lin*

Department of Physics, The Hong Kong University of Science and Technology, Clear Water Bay, Hong Kong 999077, China

†These authors contributed equally to this work
*Corresponding author: phnlin@ust.hk

The supplementary materials contain the large scale STM topographs of step (iii) and step (v) in the multi-step growth, Fast Fourier Transform (FFT) pattern of the $Ni_3(HITP)_2$ single layer, aggregation behavior of metallic nickel and $Ni(acac)_2$ on HOPG, different controlled growth experiments, evolution of coordination structures formed with $Ni(acac)_2$ with HATP molecules on Au(111) surface, thermal and chemical stability of $Ni_3(HITP)_2$ single layer, and the structure of $Ni(acac)_2$ formed on Au(111).

**Figure S1: Large scale STM topographs of step (iii) and step (v) in the multi-step growth.**

At step (iii), the Ni(acac)$_x$ clusters are 6.0-7.0 nm in size (measured by the full width at half maximum of the cluster profile) and 2.5-3.5 nm in height. At step (vi), the Ni(acac)$_x$ clusters are 5.0-5.5 nm in size (measured by the full width at half maximum of the cluster profile) and 1.5-2.0 nm in height.

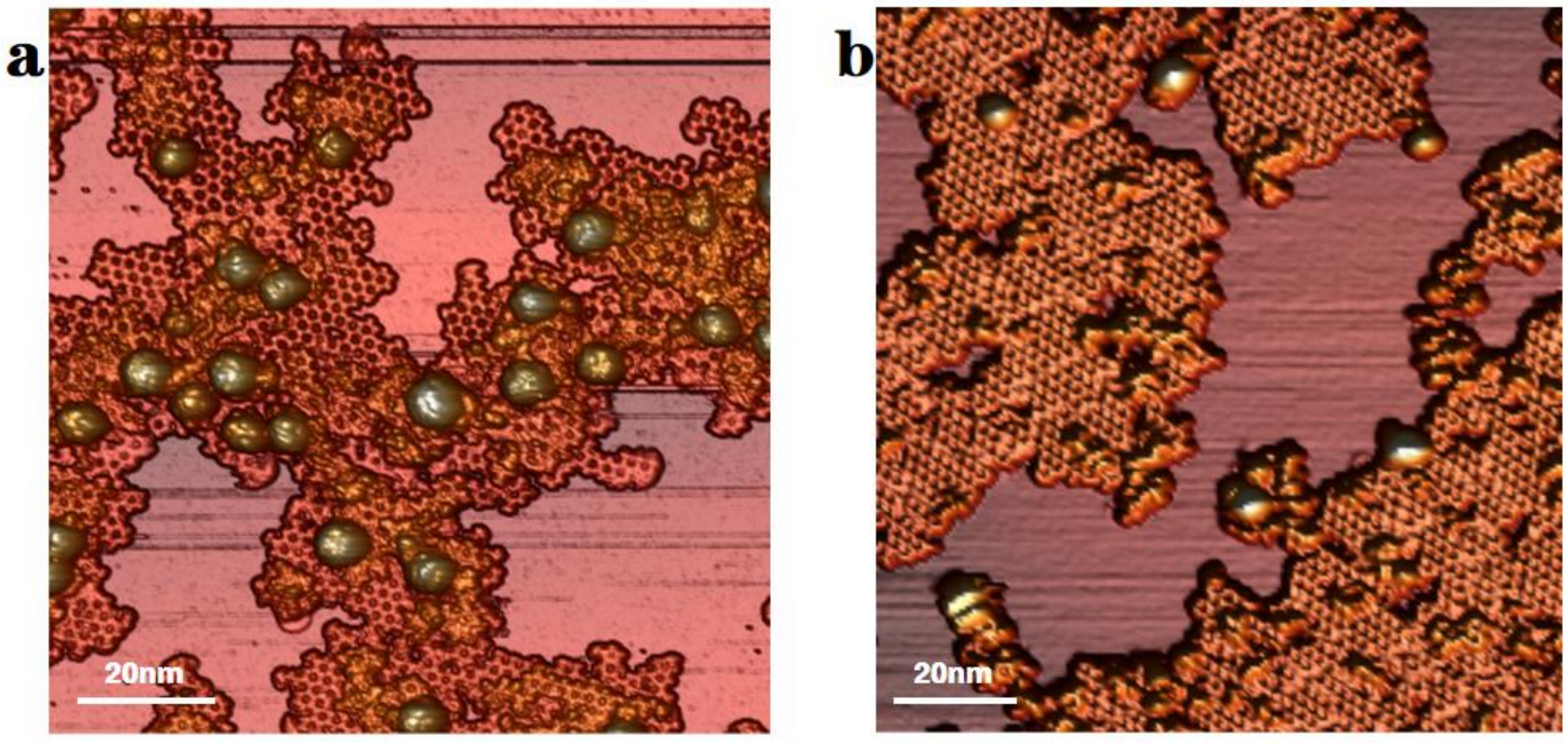

Figure S1. (a) Large scale STM topograph (3D top view) of step (iii). (b) Large scale STM topograph (3D top view) of step (v). Scanning parameters: V = −2 V, I = 500 pA.

**Figure S2: Fast Fourier Transform (FFT) pattern of the STM image showing different domains of the $Ni_3(HITP)_2$ single layer.**

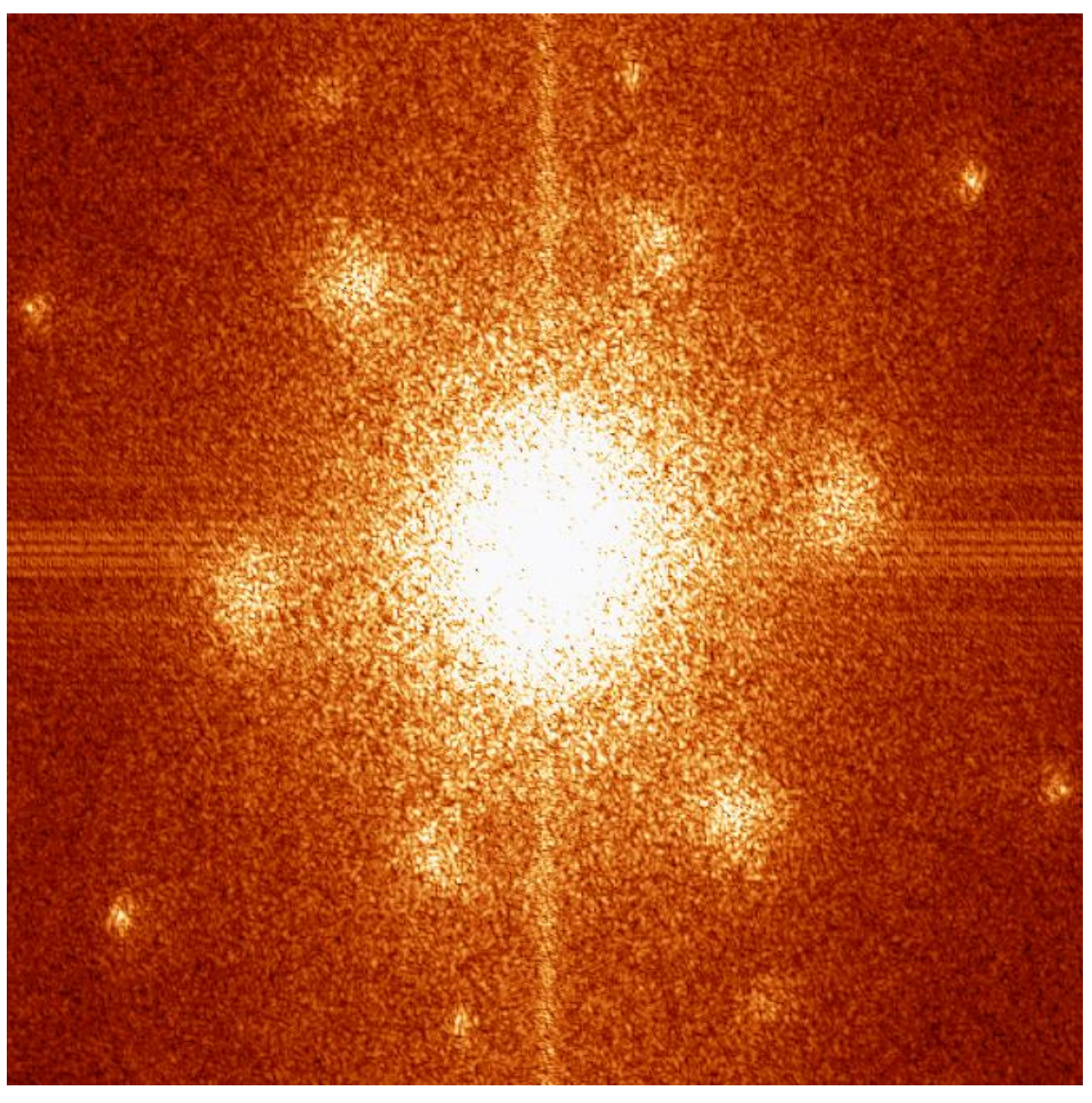

Figure S2. FFT of Figure S1(b) showing the different domains are in the same

orientation.

**Figure S3: Aggregation behavior of metallic nickel and $Ni(acac)_2$ on HOPG.**

Figure S3b shows Ni atoms aggregate into large clusters after 150 °C annealing, typically 5-7 nm in height and 8-10 nm in lateral size. These clusters predominantly adhere to step edges of the HOPG substrate. In contrast, after 150 °C annealing $Ni(acac)_2$ form smaller clusters of ~3 nm in height and ~3 nm in lateral size. These clusters aggregate on the open terraces of the HOPG substrate.

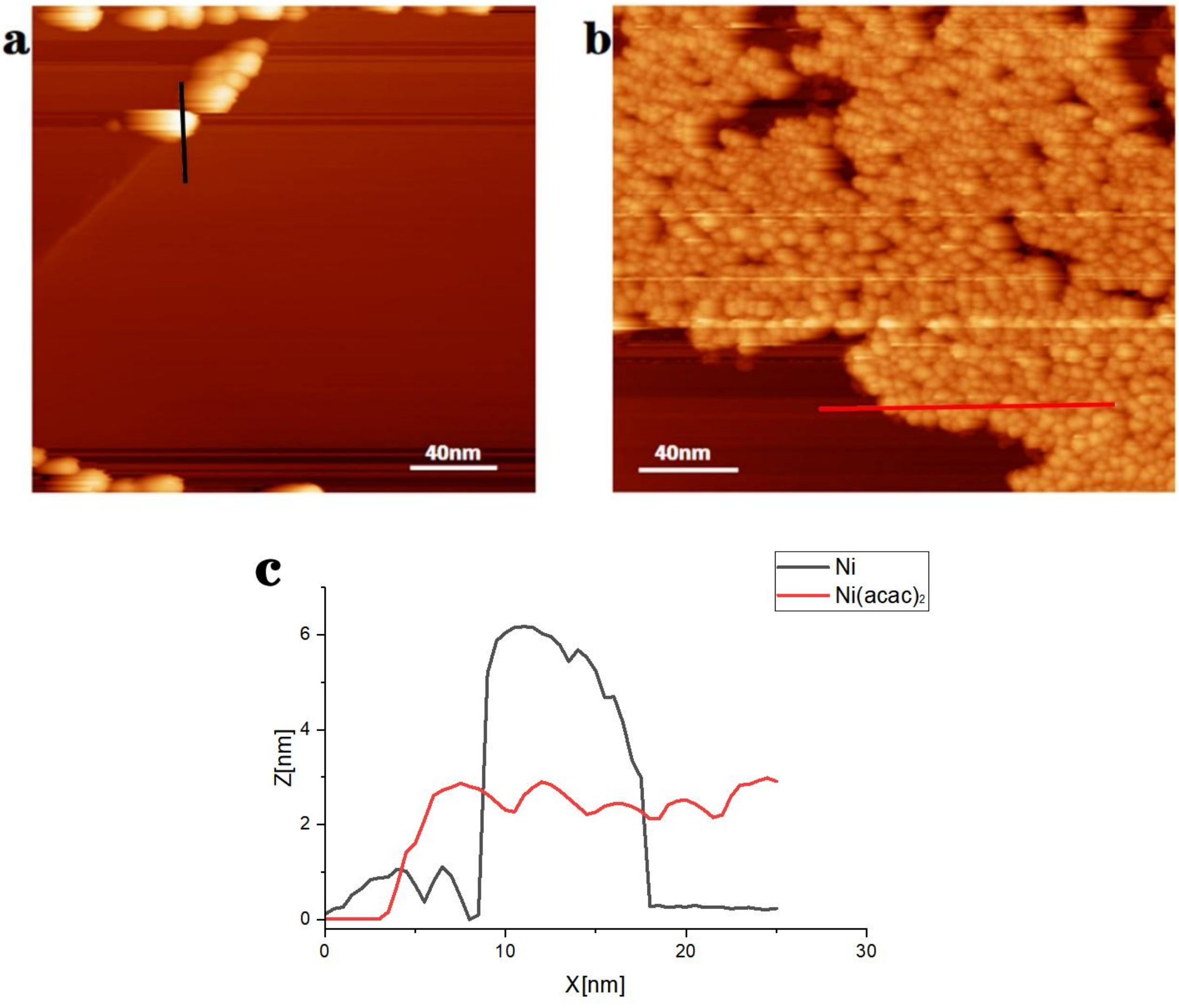

Figure S3. (a) STM topograph showing large Ni atom clusters formed on HOPG. (b) STM topograph showing small $Ni(acac)_2$ clusters formed on HOPG. (c) Height profiles of the lines indicated in (a) and (b) showing the contrast of the cluster size of the two samples. Scanning parameters: V= -2V, I= 500 pA.

**Figure S4: Co-deposition of HATP and $Ni(acac)_2$ on HOPG held at high temperature (~300 °C).**

Depositing HATP and $Ni(acac)_2$ on the hot substrate yields small pieces of $Ni_3(HITP)_2$ formed at the step edge of the HOPG. Presumably, both HATP and $Ni(acac)_2$ desorb from the hot substrate, preventing the formation of large $Ni_3(HITP)_2$ domains.

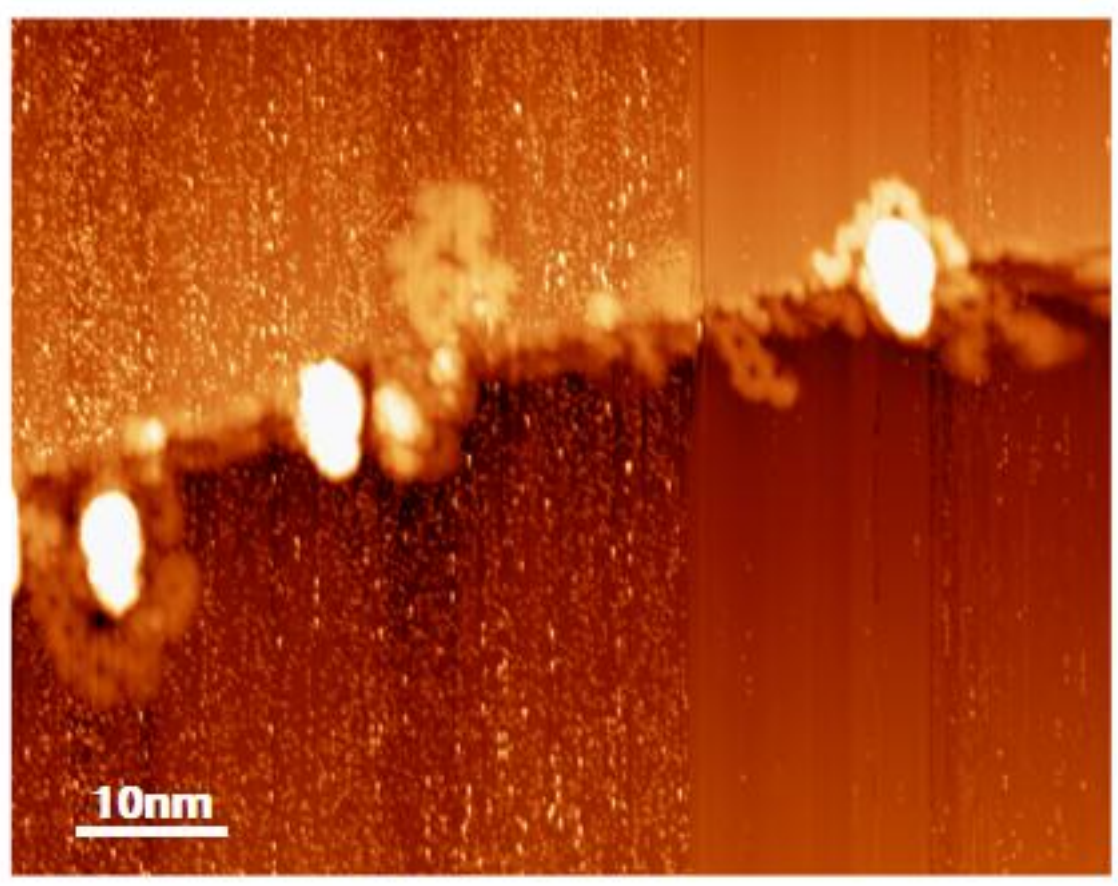

Figure S4. STM topograph showing small pieces of $Ni_3(HITP)_2$ at the step edge of the HOPG substrate formed by co-deposition of HATP and $Ni(acac)_2$ simultaneously onto a hot HOPG substrate. Scanning parameters: V= -2V, I= 500 pA.

**Figure S5: Evolution of coordination structures formed with $Ni(acac)_2$ and HATP molecules on Au(111) surface.**

Co-deposition of HATP and $Ni(acac)_2$ onto Au(111) at room temperature does not yield any ordered structures. Upon subsequent annealing (Figure S5 a–d), coordination structures are formed, suggesting Ni ions are released. At 150°C, small, short coordination oligomers such as dimers and trimers are formed. As the annealing temperature increases to 250°C, longer coordination chains and hexagonal rings are formed. At 300°C, small patches of $Ni_3(HITP)_2$ are formed. A one-step annealing at 300°C yields slightly larger $Ni_3(HITP)_2$ domains. However, these domains are much smaller than those formed on HOPG.

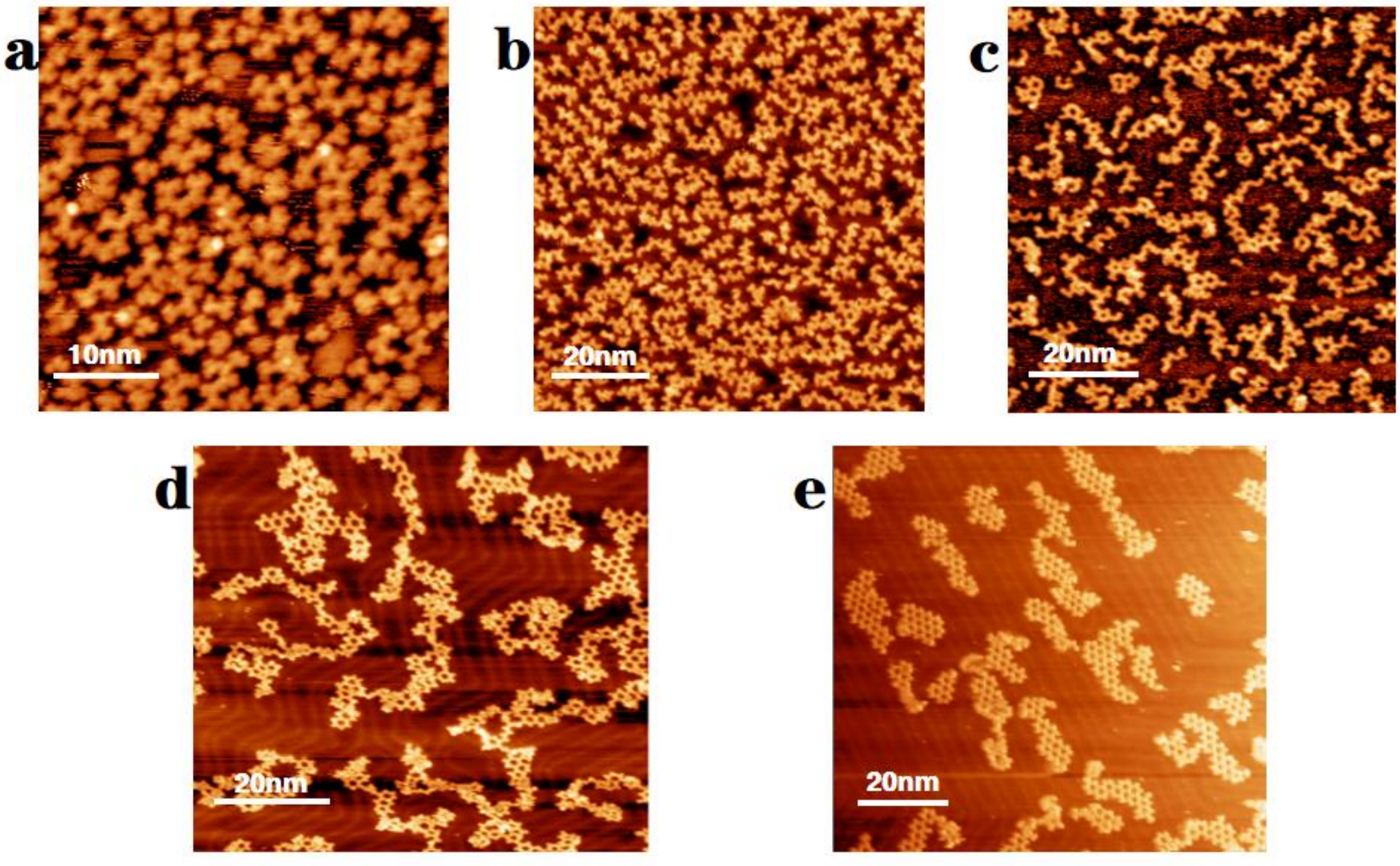

Figure S5. (a-d) Coordination structures formed with $Ni(acac)_2$ and HATP on Au(111)

subjected to stepwise annealing at different temperatures. (a) 150°C. (b) 200°C. (c) 250°C. (d) 300°C. (e): One-step annealing at 300°C without intermediate annealing steps. Scanning parameters: V= -2V, I= 500 pA.

**Figure S6: Thermal and chemical stability of $Ni_3(HITP)_2$ single layers formed on HOPG.**

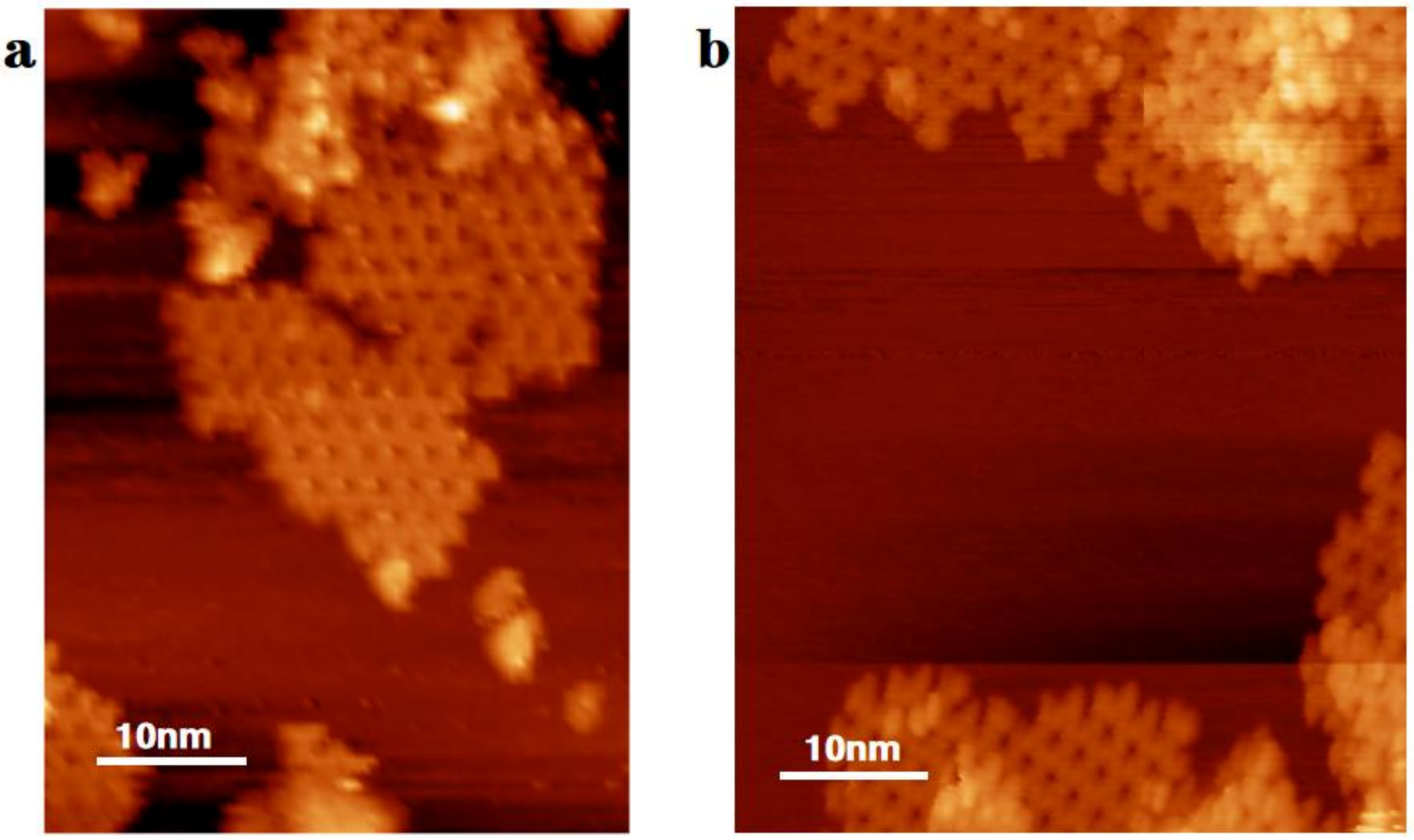

Figure S6. STM topographs showing the sample subjected to: (a) annealing at 350°C for 10 minutes. (b) exposure to ambient air for 30 minutes, re-introducing the sample back to vacuum and annealing at 150°C. Scanning parameters: V = −2 V, I = 500 pA.

**Figure S7: Structure of $Ni(acac)_2$ formed on Au(111).**

Figure S7a shows that $Ni(acac)_2$ deposited on Au(111) form closely packed one-dimensional chains. The high-resolution image of Figure 7b features that the chains are composed of periodically ranged unit with a lattice constant of 8.8 Å. This lattice constant matches the size of $Ni(acac)_2$ complex, as shown in the overlaid molecular model in Figure 7b.

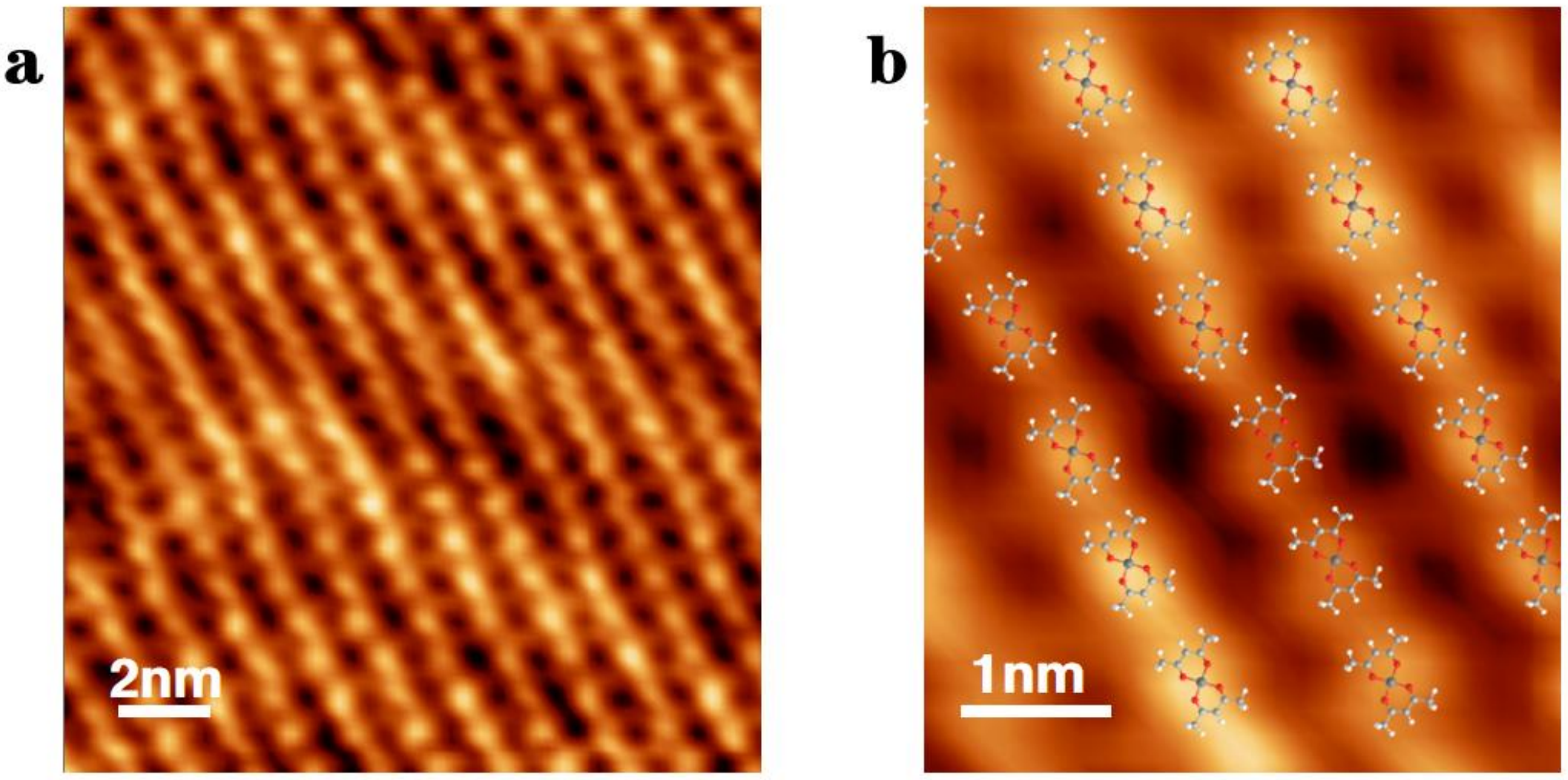

Figure S7. (a) Closely packed one-dimensional chains formed out of $Ni(acac)_2$ on Au(111) surface. (b) The periodical lattice matching the size of $Ni(acac)_2$ as illustrated by the overlaid molecular model.